# Physical Galaxy Pairs and Their Effects on Star Formation

I. M. Selim, Y.H.M. Hendy, and R. Bendary

*National Research Institute of Astronomy and Geophysics*

*(NRIAG) 11421-Helwan, Cairo, Egypt*

***i_selim@yahoo.com***

## Abstract

We present 776 truly physical galaxy pairs, 569 of them are close pairs and 208 false pairs from Karachentsev (1972) and Reduzzi & Rampazzo (1995) catalogues, which contains 1012 galaxy pairs. Also we carried out star formation activity through the far-infrared emission (FIR) in physical (truly) interacting galaxies in some galaxy pairs and compared them with projection (optical) interacting galaxy pairs. We focused on the triggering of star formation by interactions and analyzed the enhancement of star formation activity in terms of truly physical galaxy pairs. The large fraction of star formation activity is probably due to the activity in the exchange of matter between the truly companions. The star formation rate (SFR) of galaxies in truly galaxy pairs is found to be more enhanced than the apparent pairs.

## 1. Introduction

Star formation is probably the most fundamental of all astrophysical processes depends not only on the properties of stellar systems of all activity galaxies type, but also on the properties of the interaction of galaxies controlled to a large extent by various feedback effects of star formation. Studying the properties of the galaxies interaction is clear evidence for enhanced star formation rates (SFR) in interacting galaxies.

Several studies indicate that interaction between galaxies can affect their basic properties. Numerical simulations show that the star formation can be strongly affected by gravitational interaction, and some peculiarities are clearly associated with formation of bridges and tails. Simulations of merging galaxies are consistent with interaction-induced star formation, with the closest pairs (Mihos & Hernquist 1996; Montuori et al. 2010).

Galaxy pairs consisting of one physical (true) pairs and one optical (projection) pairs used to examine the effects of galaxy interaction on the star formation. A physical pair is two galaxies in real spatial, while an optical pair is two galaxies appearing close together due to only the projection effect on the sky they are not belonging to each other and totally unrelated. The interactions during close encounters could be correlated with enhancements of the star formation activity. The strongest enhancements are found in the closest pairs with projected separations 30 kpc (Barton, Geller, & Kenyon 2000; Lambas et al. 2003; Ellison et al. 2008; Freedman Woods et al. 2010).

Star formation depends on many processes and many variables, these processes and variables are interrelated in complex ways that make the subject difficult to treat in a deterministic way in the tradition of classical physics. These enhancements are accompanied by bluer central colors (Patton et al. 2011), diluted metallicities (Scudder et al. 2012), and an increased incidence of active galactic nuclei (Ellison et al. 2011) and luminous infrared galaxies (Ellison et al. 2013). In the local universe, observations showed that the merging and interacting galaxies enhance star formation processes in these galaxies and star formation activity is dependent on relative proximity and environment. (Barton, Geller, & Kenyon 2000; Patton et al. 2002; Le Fèvre et al. 200l; Lambas et al. 2003 and  Alonso et al. 2005).

## 2. Present Sample

The present sample is extracted from the catalogues with apparent companions, Karachentsev (1972), Reduzzi & Rampazzo (1995). Karachentsev built The Northern Catalogue of Isolated Pairs of Galaxies (CPG) which contains 603 binary galaxies. Reduzzi & Rampazzo (1995) used the Karachentsev criteria and applied it for ESO-Uppsala catalogue of the southern hemisphere (Lauberts & Valentijn 1989: ESO-LV) and obtained 409 isolated galaxy pairs. Accordingly, the sample of 1012 isolated binary galaxies is considered as a homogeneous sample for our study.

## 3. Galaxy Pairs Classifications

For every galaxy pairs mentioned in the above catalogues, we identified the closest pair due to the smallest true physical separation. We used the criteria ($Cr$) of Ali (2000) and Lambas et al. (2003) with the difference in velocity ($\Delta V$) and projected separation ($rp$), to obtain the results shown in table 1. The value of $Cr = 390$ was found by Ali (2000) to be the statistical discriminator between true pairs and false one. Therefore, pairs with $Cr \leq 390$ are considered to be true galaxy pairs, while the pairs having $Cr > 390$ are considered to be false pairs. Physical galaxy pairs with $Cr \leq 130$ to be considered a close pairs, Selim et al. (2007).

Our sample contains 984 galaxy pairs with available data (separation between pair members in arcminutes, radial velocity (km/s), semimajor (a), semiminor (b) axes in arcminutes of each galaxy in the galaxy pairs). 28 galaxy pairs had not available data. 776 galaxy pairs are true galaxy pairs (~ 79%), 208 galaxy pairs are false galaxy pairs (~ 21%) Table.1.

Table. 1

| Catalogue | All pairs | Close pairs | Normal pairs | False pairs | Unstudied |
|---|---|---|---|---|---|
| Karachentsev (1972) | 603 | 328 | 171 | 86 | 18 |
| Reduzzi & Rampazzo (1995) | 409 | 241 | 36 | 122 | 10 |
| Criterion | - | $Cr \leq 130$ | $Cr \leq 390$ | $Cr > 390$ | - |

## 4. SFR Enhancement as a Function of Separation

When a galaxy interacts with another galaxy, its disk becomes gravitationally unstable. This instability induces a non-axisymmetric distortion such as a bar, which drives gas of the disk into the circumnuclear region, Lehnert & Heckman (1996). The final true galaxy pairs allow a detailed study of the possible effects of the interactions on the star formation rate. So we have estimated (SFR) for true galaxy pairs in Karachentsev catalogue (1972) and Reduzzi & Rampazzo catalogue (1995). The FIR-luminosity has been used to estimate the SFR.

SFR derived from the FIR luminosity was given by Thronson, et al. (1989) as follows:

$$SFR_{FIR}\,(M_{\odot}\,yr^{-1}) = 6.5 * 10^{-10} * L_{FIR}\,(L_{\odot}),$$

$L_{FIR}$ ($L_{\odot}$) is the FIR luminosity from the *IRAS* observation in 60 μm and 100 μm bands and is estimated from the following equation (Thronson, et al.1991),

$$L_{FIR}\,(L_{\odot}) = 6 * 10^{5}\,(2.58\, S_{60} + S_{100})\ D^{2},$$

Where D is the distance in Mpc, *S60* and $S_{100}$ are the IRAS flux densities in Jy at 60 and 100μm respectively. The values of SFR are estimated for some galaxies in Karachentsev catalogue (1972). Reduzzi & Rampazzo catalogue (1995) using the IRAS observation in 60 and 100 μm. We have used a well-defined sample of star-forming galaxies to measure SFR enhancements as a function of pair separation. We found that the SFR enhancements are detectable out of *rp* ~ 300 kpc, with no enhancement seen at larger galaxies separations. Also, we found that the star formation as a function of *Cr* increases for closer galaxy pairs than the all galaxy pairs in Reduzzi & Rampazzo catalogue (1995), see (Fig.1). As with earlier work Selim et al. (2007) found the strongest enhancements at the smallest pair separations. A similar behavior is found in catalogue of Karachentsev (1972) see Figs. 1, 2, 3 and 4 showing the star formation rate as a function of *rp* and *Cr.* A negative weak correlation is found between the reliable upper limits of the galaxy pairs for the *rp, ΔV* and *Cr* with SFR for the true pairs.

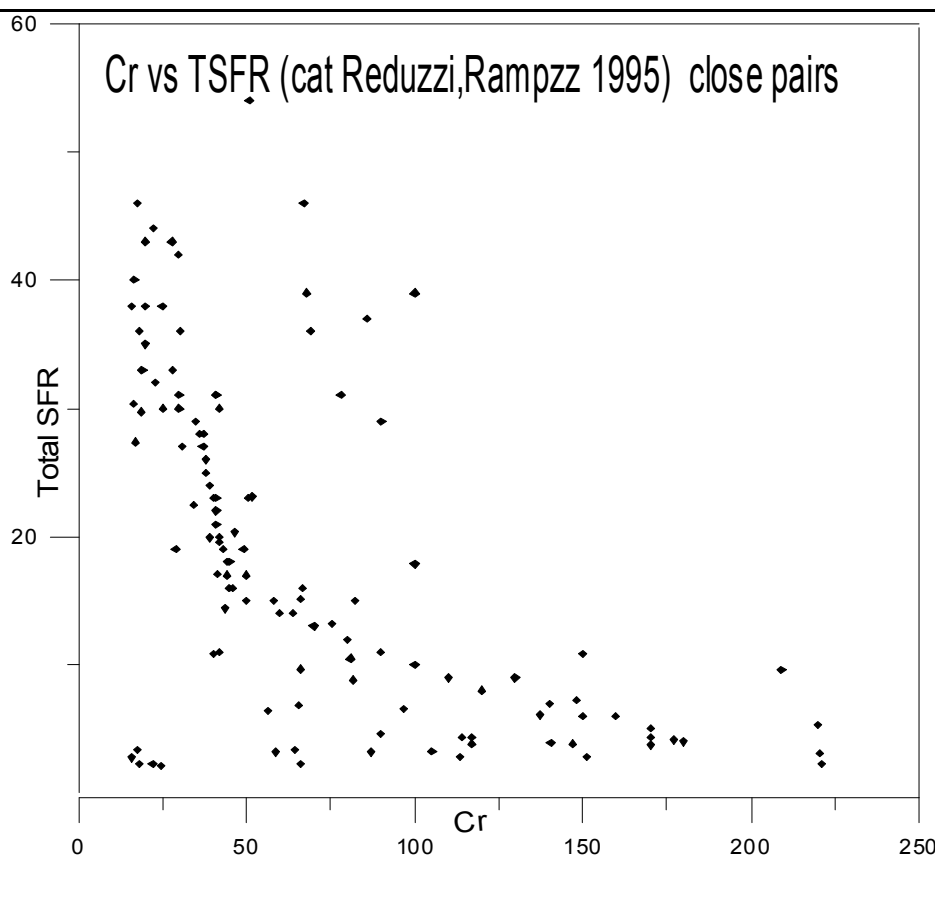

(a) *Cr* vs SFR (catalogue of Reduzzi and Rampazzo (1995)) for all pairs.

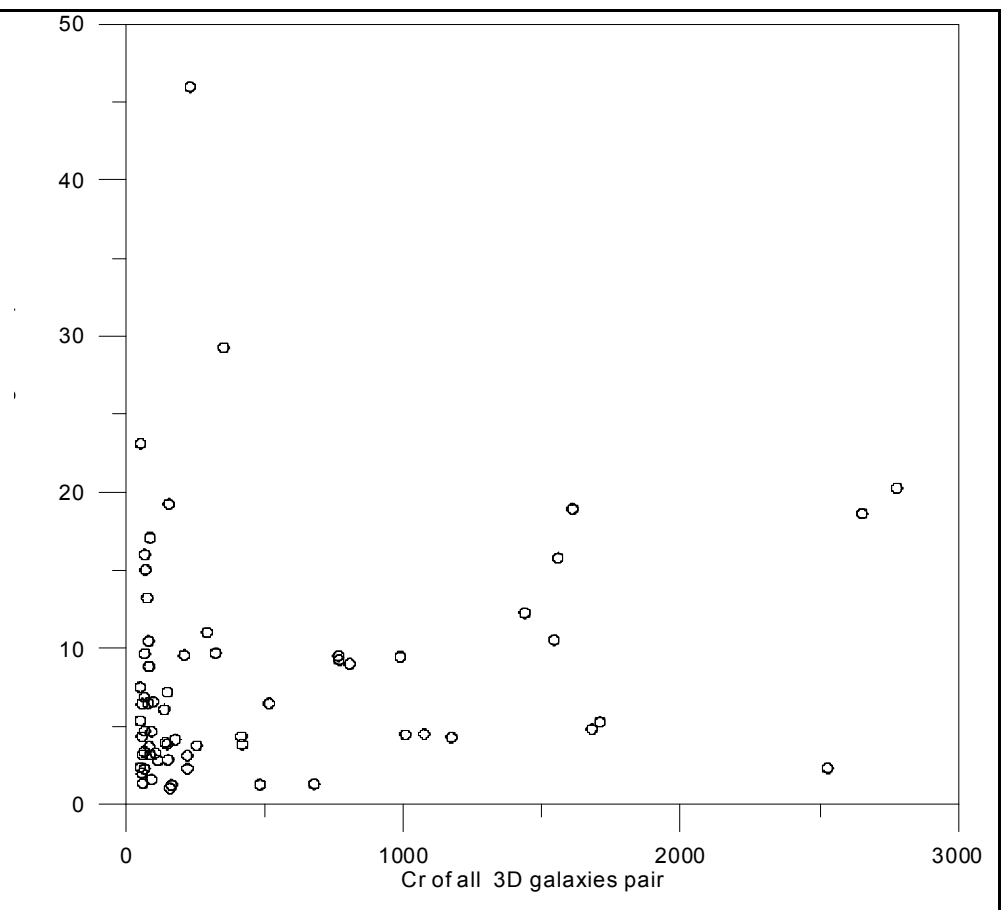

(b) *Cr* vs SFR (catalogue of Reduzzi and Rampazzo (1995)) for close pairs.

Figure 1. (a): The star formation rate as a function of *Cr*-values for all galaxy pairs and (b): for close pairs of catalogue of Reduzzi & Rampazzo (1995).

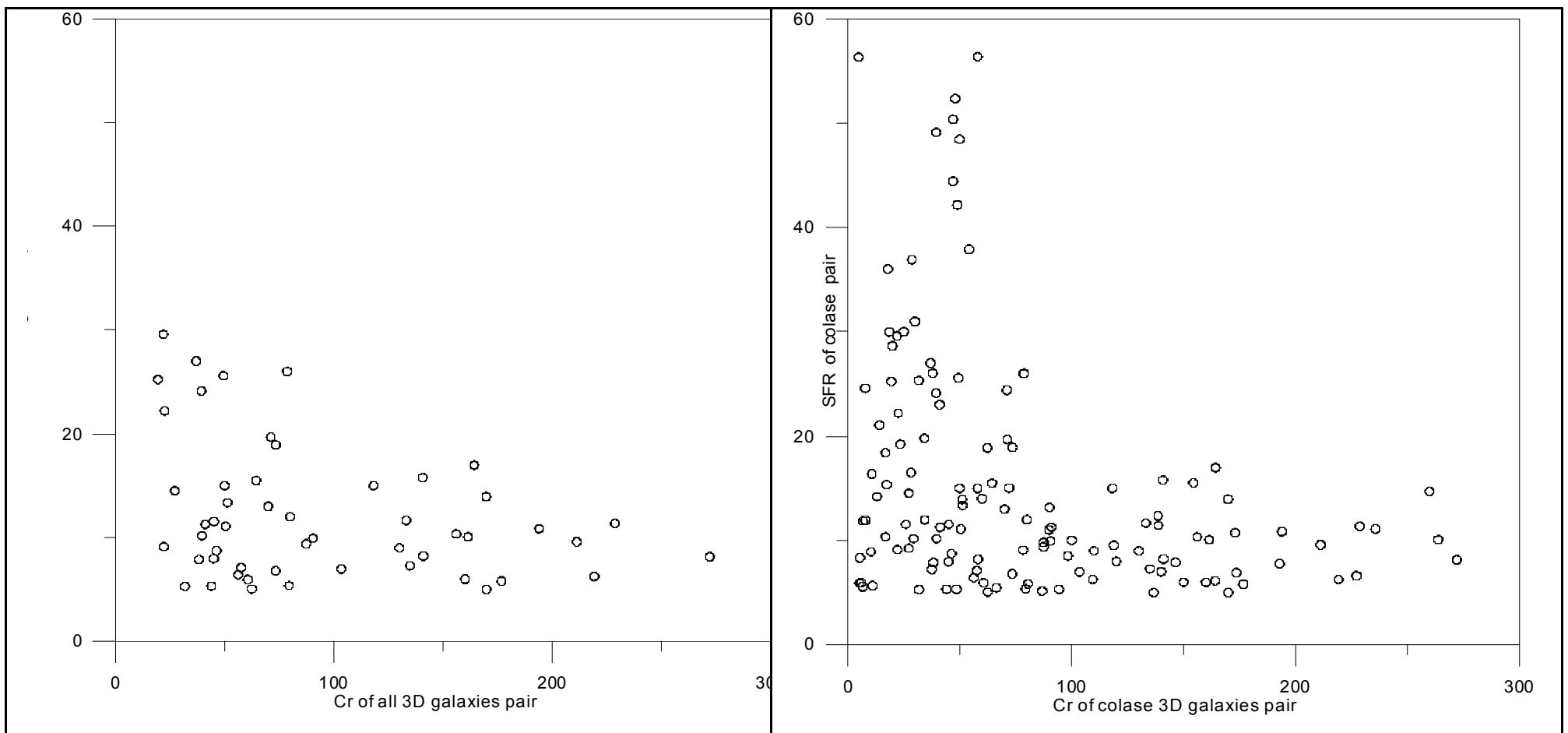

(a) *Cr* vs SFR (Karachentsev catalogue 1972) for all pairs.

(b) *Cr* vs SFR (Karachentsev catalogue 1972) for close pairs.

Figure 2. The star formation rate against *Cr* of Karachentsev catalogue.

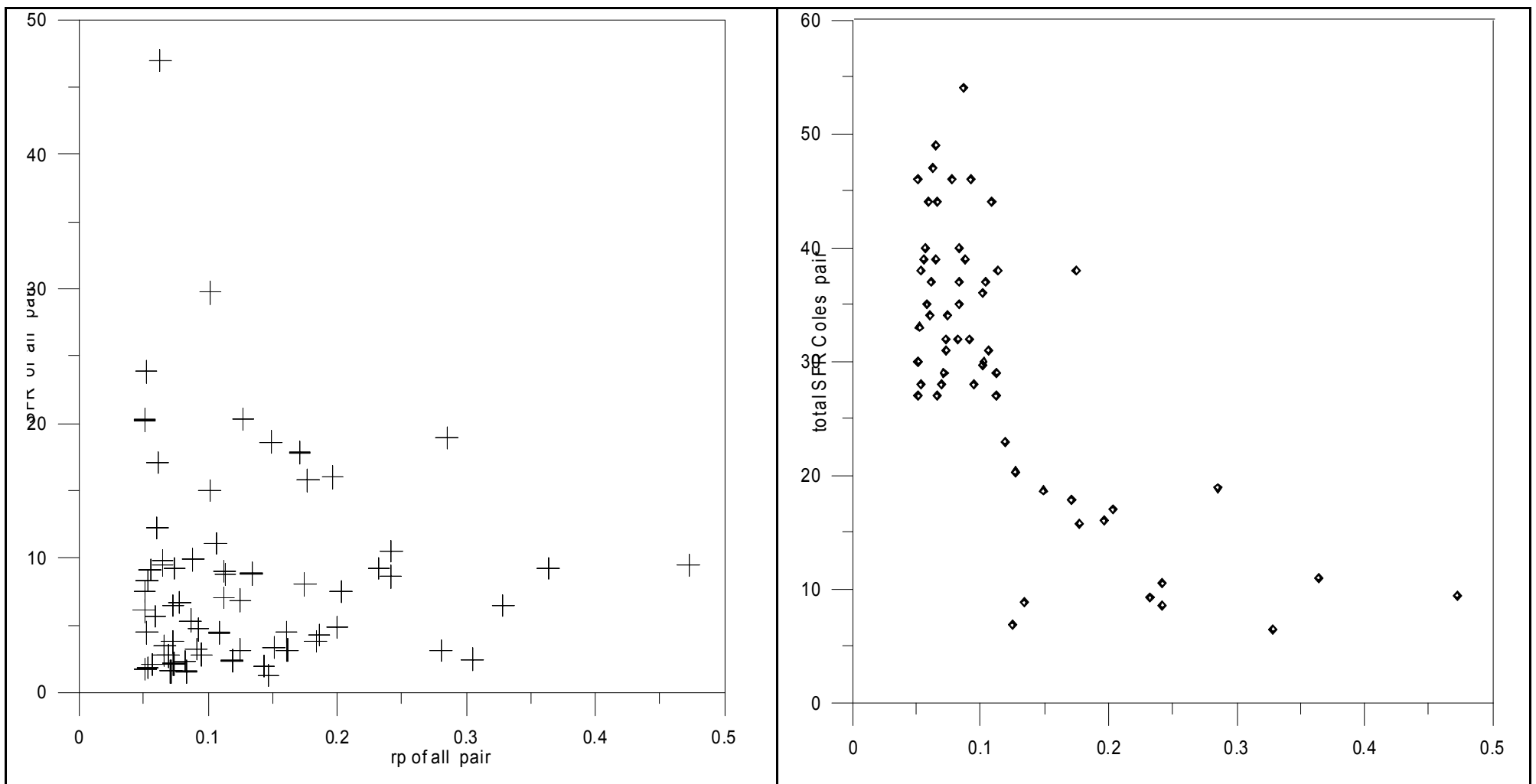

(a) *rp* vs SFR of all pairs (catalogue of Reduzzi & Rampazzo (1995)).

(b) *rp* vs SFR of of close pairs (catalogue of Reduzzi & Rampazzo (1995)).

Figure 3. The star formation rate against *rp* (catalogue of Reduzzi & Rampazzo (1995)).

.

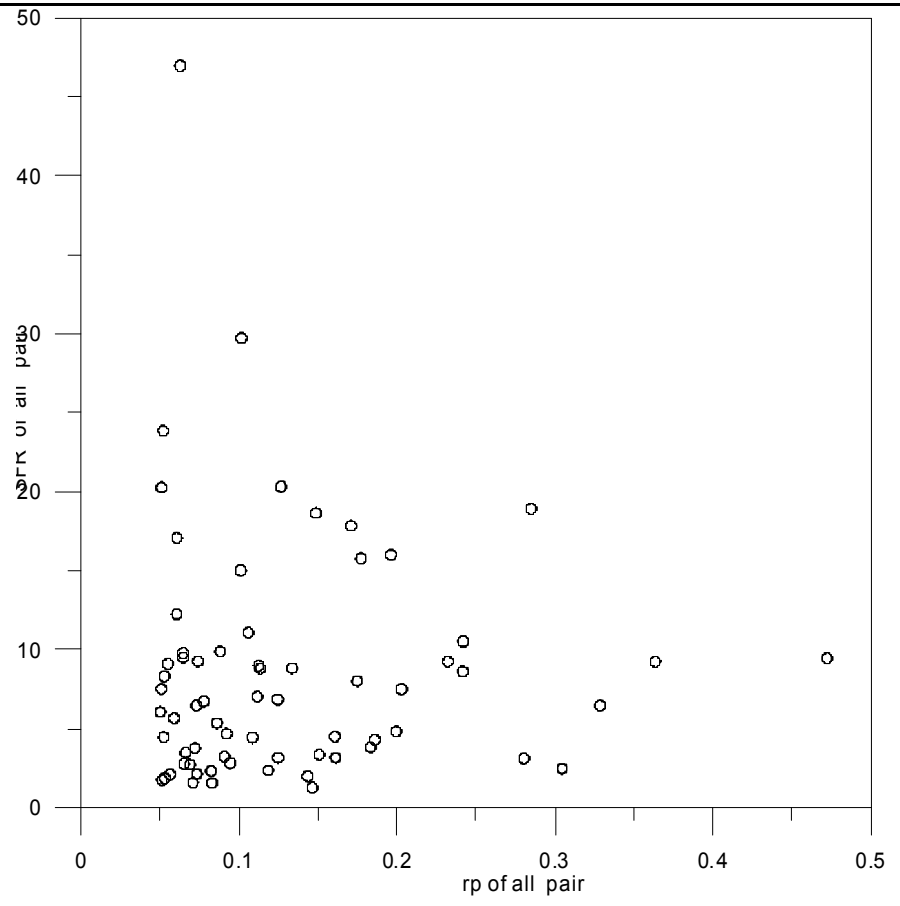

(a) rp vs SFR of all pairs (Karachentsev catalogue (1972)).

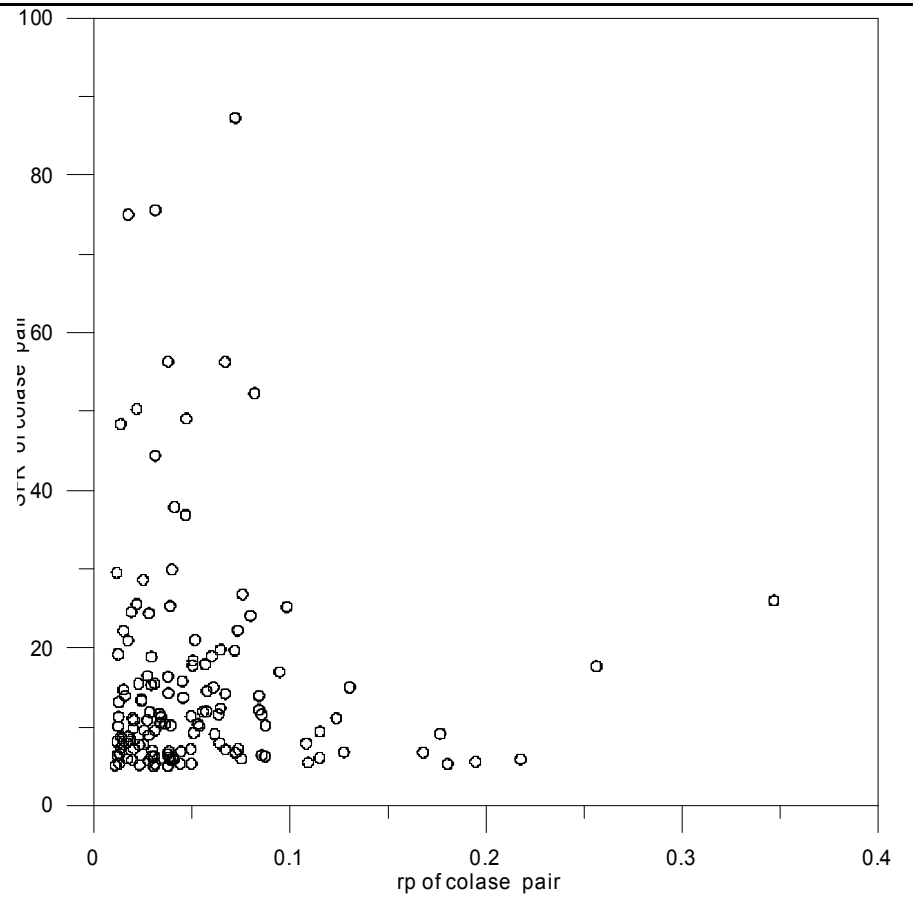

(b) rp vs. SFR of close pairs (Karachentsev catalogue (1972)).

Figure 4. The star formation rate against *rp* (Karachentsev catalogue 1972).

## 5. Discussion and Conclusion

To investigate the dependence of specific star formation rate on true galaxy pairs, we considered firstly the dependence of SFR on *rp* separation, looking subsequently at the dependence on $\Delta V$. We investigated 1012 galaxy pairs, (776 of them are truly physical galaxy pairs, 208 of them are false pairs Table. 1). Also we analyzed the dependence of star formation on type of galaxy pairs by defining the usual projected galaxy pairs and physical galaxy pairs.

Galaxy pairs show weak correlation between star formation activity and proximity between the components. The star formation rate in close pairs is similar to the results of Barton et al. (2000) and Lambas et al. (2003). We found a good agreement with the results of previous studies of galaxies interaction, (Selim et al. 2007; Plana et al. 1998, 2003; Allam et al. 1996; and Amram et el. 2004) for interactions to statistically enhance star formation activity. We conclude that the star formation rate appears to be a weaker function of projected physical separation than the velocity difference. Close galaxy pairs show significant correlation between star formation activity and proximity to a companion. These indicate that the SFR enhancements at small separations are due to systems near the end of the merger process. There is also a trend showing the increase of the star formation activity with proximity between the components.

## Acknowledgements

The authors would like to thank the Associate Professor M.R. Sanad
for the useful and valuable comments of this paper.